\documentclass[aps,prl,reprint,groupedaddress]{revtex4-1}

\usepackage{amsmath}
\usepackage{graphicx}
\begin{document}


\title{Space charge fields in azimuthally symmetric beams: integrated Green's
  function approach}


\author{Petr M.\ Anisimov}
\email[]{petr@lanl.gov}
\affiliation{Los Alamos National Laboratory, Los Alamos, New Mexico, 87545,
  USA}

\author{Nikolai A.\ Yampolsky}
\affiliation{Los Alamos National Laboratory, Los Alamos, New Mexico, 87545,
  USA}

\date{\today}

\begin{abstract}
Electromagnetic fields induced by the space charge in relativistic beams play 
an important role in Accelerator Physics. They lead to emittance growth, slice 
energy change, and the microbunching instability. Typically, these effects are 
modeled numerically since simple description exists only in the limits of
large- or small-scale current variations. In this paper we consider an axially
symmetric charged beam inside a round pipe and find the solution of the space
charge problem that is valid in the full range of current variations. We
express the solution for the field components in terms of  Green's
functions, which are fully determined by just a single function. We then find
that this function is an on-axis potential from a charged disk in a round
pipe, with transverse charge density $\rho_{\perp}(r)$, and
it has a compact analythical expression. We finally provide an
integrated Green's function based approach for efficient numerical evaluation
in the case when the transverse charge density stays the same along the beam. 
\end{abstract}

\maketitle

\section{Introduction}

The space charge effect is a basic collective phenomenon in Accelerator
Physics that plays an important role both in electron and proton machines. In
high current regimes, self-generated electromagnetic fields become so strong
that they lead to significant emittance growth
\cite{a:Chasman:1969,a:Lapostolle:1971,a:Hofmann:1981,a:Kim:1989}, slice
energy change, and the microbunching instability \cite{a:Huang:2004}, which is
especially detrimental in the context of free electron laser linacs
\cite{a:Borland:2002,a:Huang:2004}. In the case of x-ray free electron lasers
with periodic enhancement of electron peak current
\cite{a:Zholents:2005,a:Kur:2011,a:Tanaka:2013}, strong longitudinal current
modulation may result in large space charge forces
which, in turn, may limit the performance of these schemes. 

Compact analytical expressions for the space charge induced fields are
currently available in the limits of either large- or small-scale current
variations. The first limit provides a local description \cite{b:Chao:1993}
while the second limit requires an impedance based description
\cite{tr:Rosenzweig:1996} that is non-local as it depends on the Fourier
spectrum of the current. This latter approach allows for semi-analytical
treatment and is implemented in the Elegant numerical code
\cite{tr:Borland:2000} in order to include longitudinal space charge effects
on a beam axis.  

An alternative approach adopted in many numerical codes, such as OPAL 
\cite{tr:OPAL:2008}, Astra \cite{web:Floettmann:2011}, and Parmela
\cite{tr:Billen:1996}. This approach uses a Poison solver in order to find the
space charge induced fields. However it is very time consuming and a
semi-analytic description for the space charge induced fields applicable in
the full range of current variations is highly desired.  

In this paper we provide a semi-analytical description of the space charge
problem in the case of an axially symmetric beam in a round pipe. The derived
expressions for components of the induced field are valid in the full range of
current variations. They also can be efficiently evaluated by the method of
Integrated Green's functions (IGF). 

We present our approach in Section \ref{sec:II}, where we find that the
Green's function for a charged disk in a round pipe fully determines the
components of the induced fields. In Section \ref{sec:GF-analysis}, we
express the Green's function in terms of an on-axis potential from a charged
disk in a round pipe and find a compact analytical approximation for this
potential. Section \ref{sec:IGF-approach} uses the compact analytical
expression for the Green's function in order to present the field components
in a form suitable for the IGF approach. In Section
\ref{sec:IGF-implementation} we suggest how to improve a semi-analytical
description of the space charge fields in  numerical codes such as Elegant by
providing a step-by-step instructions of IGF approach with our Green's
function. We finally summarize our findings in the conclusion.  

\section{\label{sec:II} Basic Equations for the Space Charge Fields}

\begin{figure}[t]
  \centering
  \includegraphics[width=82mm]{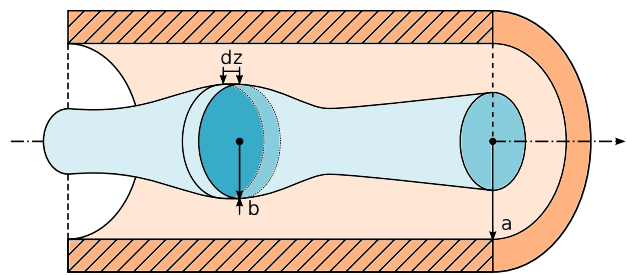}
  \caption{An axially symmetric electron beam in a perfectly
    conducting pipe of radius $a$. A disk section of length $dz$ and the
    transverese charge density $\rho_{\perp}\left( r , z
    \right)$, where $z$ is a coordinate along the beam in the beam reference
    frame, is shown.}
  \label{fig:model}
\end{figure}

The most general approach considers an electron beam with charge density
$\rho(\mathbf{r})$ that travels on axis of a perfectly conducting pipe of
radius $a$ at a speed $v_{z}$ (see Fig.~\ref{fig:model}). It starts in the
beam frame, $\mathbf{R}$, where the electrostatic potential induced by the
space charge, $\varrho(\mathbf{R})$, is a solution of the Poisson equation:
\begin{equation}
  \label{eq:general-potential}
  \Phi(\mathbf{R}) = \frac{1}{4\pi\varepsilon_{0}} \int \varrho(\mathbf{R'})
  \Gamma(\mathbf{R}|\mathbf{R'}) d^{3}\mathbf{R'},
\end{equation}
where $\Gamma(\mathbf{R}|\mathbf{R'})$ is the Green's function for the
Laplace's equation inside a perfectly conducting pipe. The corresponding
Green's function is derived in 
Appendix as an expansion in terms of radial-azimuthal eignenfunctions that can
be used in systems without translational symmetry.

In this paper, we will consider axially symmetric densities in the lab frame,
$\mathbf{r}$, of the 
form $\rho(\mathbf{r})=\rho_{\perp}(r)\lambda(z)$ with normalization $\int
\rho_{\perp}(r')d^{2}\mathbf{r'}=1$. In this case, the electrostatic potential
in the beam frame is expressed as
\begin{equation}
  \label{eq:symmetric-potential}
  \Phi(R,Z) = \frac{1}{4\pi\varepsilon_{0}} \int_{-\infty}^{\infty}\Lambda(Z')
  G(R,Z-Z') dZ',
\end{equation}
where $\Lambda(Z')=\gamma^{-1}\lambda(Z'/\gamma)$ and
\begin{equation}
  \label{eq:axial-Greens}
  G(R,Z-Z') = \sum_{n=1}^{\infty} \frac{2 c_{n} J_{0}(\mu_{0,n}R/a)}{a
    \mu_{0,n}J_{1}(\mu_{0,n})^{2}}    e^{-\mu_{0,n}|Z-Z'|/a},
\end{equation}
with $c_{n} = \int \rho_{\perp}(r') J_{0}(\mu_{0,n}r'/a) d^{2}\mathbf{r'}$
that has to be individually calculated for different transverse density profiles. In
the particular case of $c_{n}=1$, one recovers a result for an electrostatic
potential of a point charge on the axis of a perfectly conducting pipe (see
Eq.~(27) in Ref.~\cite{a:Bouwkamp:1947}). 

The electrostatic potential in the beam reference frame results in the
electric field with longitudinal and transverse components that have the
following expressions in the lab frame
\begin{equation}
  \label{eq:Ez}
E_{z}(r,z)=-\frac{1}{4 \pi \varepsilon_{0} \gamma} \int _{-\infty }^{\infty}
\lambda(z')
\frac{\partial}{\partial z} g(r, z-z') dz',
\end{equation}
\begin{equation}
  \label{eq:Er}
 E_{r}(r,z)=-\frac{\gamma}{4 \pi \varepsilon_{0} } \int _{-\infty }^{\infty }
 \lambda(z')
 \frac{\partial}{\partial r} g(r, z - z') dz',  
\end{equation}
in terms of transformed to the lab frame Green's function,
$g(r,z-z')=G(r,\gamma z-\gamma z')$.
Additionally, in the lab frame, there is also an azimuthal magnetic field
$B_{\phi}(r,z)=\frac{v_{z}}{c^{2}}E_{r}(r,z)$, which reduces the overall
transverse space charge force acting on charged particles in the beam by
$\gamma^{-2}$.

\section{\label{sec:GF-analysis} Green's function analysis}

In the previous section, we have reduced the problem of space charge induced
fields to the finding Green's function, $G(R,Z-Z')$, --- the electrostatic
potential from a charge distribution, $\varrho( \mathbf{R} ) = \rho_{\perp}(R)
\delta(Z-Z')$. Due to the axial symmetry of the charge distribution, one can
use the Poisson representation \cite{a:Poisson:1823} in order to the potential
outside the axis of symmetry once the potential on the axis has been found: 
\begin{equation}
  \label{eq:potential_average}
  G(R,Z-Z') = \frac{1}{2\pi} \int_{0}^{2\pi} G(0, Z - Z' + i R \cos \phi) d\phi,\ Z>Z', 
\end{equation}
where
\begin{equation}
  \label{eq:on-axis-Greens}
  G(0,Z-Z') = \sum_{n=1}^{\infty} \frac{2 c_{n}}{a
    \mu_{0,n}J_{1}(\mu_{0,n})^{2}}   e^{- \mu_{0,n}|Z-Z'|/a}.
\end{equation}
Sometimes, however, the integral cannot be taken analytically or potential has
to be found near the axis. Therefore, we propose an alternative approach here
based on the formal separation of variables: 
\begin{equation}
  \label{eq:op-response-func}
G(R,Z-Z') \equiv J_0\left(R \frac{\partial}{\partial Z}
\right)G(0,Z-Z'),\ Z\ne Z',
\end{equation}
where a function of a derivative is defined via its Taylor series expansion
\footnote{Consider the following on axis potential, $G(0,Z-Z')=1/|Z-Z'|$. 
  Using the Poisson representation, one can show that
  $G(R,Z-Z')=1/\sqrt{(Z-Z')^{2}+R^{2}}$, which is a free space potential
  of a point charge in cylindrical coordinates. In order to apply our
  approach, one start with the Taylor series expansion, $J_{0}(x) = 1 -
  \frac{x^{2}}{4} +  O(x^{4})$, and arrives to $G(R,Z-Z') = 1/|Z-Z'|
  -\frac{R^{2}}{4} \frac{\partial^{2}} {\partial Z^{2}} 1/|Z-Z'| +
  O(R^{4})$. Carrying out the derivatives leads to $G(R,Z-Z') = 1/|Z-Z'|
  \left(1 -\frac{R^{2}}{2(Z-Z')^{2}} + O(R^{4})\right) $, which is a series
  expansion of a free space potential of a point charge in
  cylindrical coordinates for small $R$. }.
As a consequence of our approach, one can confirm, based on a formal
substitution $x \to R \frac{\partial}{\partial Z}$ into the Bessel
differential equation, $\left(x^{2}\frac{d^{2}}{dx^{2}}+x\frac{d}{dx}+x^{2} \right)
J_{0}\left(x\right)=0$, that the derived Green's function is indeed a solution of the
Laplace equation, $\left(\frac{\partial^{2}}{\partial R^{2}} +\frac{1}{R}
  \frac{\partial}{\partial R}+\frac{\partial^{2}}{\partial Z^{2}} \right)
G(R,Z-Z')=0$.  

\begin{figure}[t]
  \centering
  \includegraphics[width=82mm]{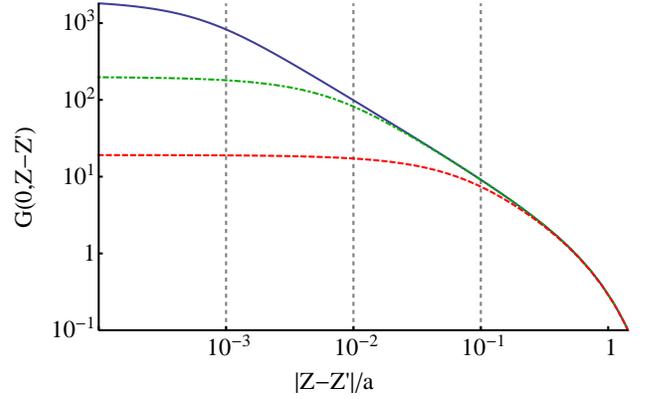}
   \caption{(Color online) An on-axis Green's function, $ G(0,Z-Z')$, in case
     of uniform transverse distribution for different values of $b/a =$ 0.1
     (dashed red), 0.01 (dot dashed green) and 0.001 (solid blue). The
     short-range behavior, $|Z-Z'|\ll a$, corresponds to the potential from a
     uniformly charged disk of radius $b$ in free space.  As distance
     increases, $|Z-Z'|\gg b$, the on-axis Green's function becomes the
     potential of a point charge in a pipe. Transformation to the lab frame
     substitutes $Z \to \gamma z$ thus scaling the transitions points to
     $a/\gamma$ and $b/\gamma$ correspondingly.
    } 
  \label{fig:F}
\end{figure}

As the transformed Green's function is $g(r,z-z')=G(r,\gamma z- \gamma z')$,
we have just reduced the space charge problem to finding electrostatic 
potential on the pipe axis, $G(0,Z-Z')$, which is shown in Fig.~\ref{fig:F},
in the case of a uniform transverse distribution, $\rho_{\perp}(r<b)=1/\pi
b^{2}$, for different values of $b/a$. It shows that there are two distinct
regions with different behaviors of the Green's function. The first region
corresponds to the far zone, $|Z-Z'|\gg b$, where the Green's function is just
a potential of a point charge in a pipe. The second region corresponds to the
vicinity of the source, $|Z-Z'|\ll a$, in which case the free space
approximation can be used. Both of these regions allow for accurate and
compact analytical expressions for $G(0,Z-Z')$. Furthermore, we can be
combined these expressions together in a single analytical expression due to a
broad overlap region, $b \ll |Z-Z'| \ll a$, which corresponds to a point
charge in a free space. 

\begin{figure}[t]
  \centering
  \includegraphics[width=82mm]{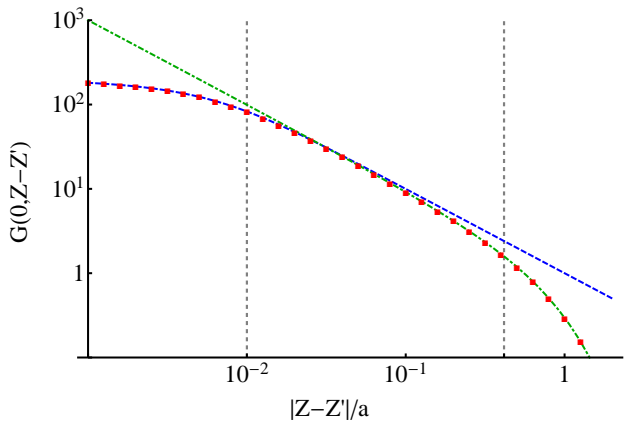}
  \caption{(Color online) The on-axis Green's function, $
    G(0,z-z')$ in the case of uniform transverse charge density with
    $b/a=0.01$ (red squares). The near (dashed blue) and far
    (dot-dashed green) zone contributions are plotted separately. The boundaries
    of these zones $|Z-Z'|=\mu_{0,1}^{-1}a$ and $|Z-Z'|=b$ are marked by the
    vertical dashed lines. Transformation to the lab frame
     substitutes $Z \to \gamma z$ thus scaling the transitions points to
     $\mu_{0,1}^{-1}a/\gamma$ and $b/\gamma$ correspondingly.}
  \label{fig:Fs-Fi}
\end{figure}

In the first region, $|Z-Z'|\gg b$, the sum in Eq.~(\ref{eq:on-axis-Greens})
converges rapidly due to vanishing exponential factors
$\exp(-\mu_{0,n}|Z-Z'|/a)$. The actual number of terms contributing to the sum
can be estimated as $\mu_{0,n}|Z-Z'|/a \sim 1$. Under this condition,
$c_{n}\simeq 1$, resulting in the expression for the on-axis potential from a
point charge on the pipe axis \cite{a:Bouwkamp:1947}. A compact form
representation for the on-axis Green's function in the far zone is hence a sum
of the geometric series:
\begin{equation}
  \label{eq:core_sum}
 G(0,Z-Z') \approx \frac{1}{ a}
 \frac{\pi e^{-\mu _{0,1}|Z-Z'|/a}}{1-e^{-\pi |Z-Z'|/a}},\ |Z-Z'|\gg b,
\end{equation}
where we have used that $\mu_{0,n}-\mu_{0,1}\approx\pi(n-1)$ and
$\mu_{0,n}J_{1}(\mu_{0,n})^{2} \approx 2/\pi$. 

The second region is in the vicinity of the source, $|Z-Z'| \ll a$. Hence, one
can ignore a boundary effect of the pipe surface but has to account for an
actual transverse charge density, $\rho_{\perp}(r)$. In the absence of the
boundary effect the convergence of the  sum in Eq.~(\ref{eq:on-axis-Greens}) is
defined by $c_{n}$ rather than by the exponent. Hence, one can replace
summation with integration. We chose the integration variable to be
$x=\mu_{0,n}$  with the Jacobian of this transformation $dn/dx\simeq 1/\pi $:
\begin{equation}
  \label{eq:short-range-Greens}
  G(0,Z-Z') \approx \frac{1}{a}\int d^{2}\mathbf{r'} \rho_{\perp}(r') \int_{0}^{\infty}
  J_{0}(x r'/a)  e^{-x |Z-Z'|/a} dx.
\end{equation}
Carrying out one integration results in the following compact form
representation of the on-axis Green's function 
\begin{equation}
  \label{eq:core_int}
G(0,Z-Z') \approx \int d^{2}\mathbf{r'}
\frac{\rho_{\perp}(r')}{\sqrt{(Z-Z')^{2}+r'^{2}}}, \ |Z-Z'|\ll a, 
\end{equation}
which is indeed the on-axis potential from a charged disk in free space. The
Table \ref{tab:transverse-distributions} provides examples of this potential
for different transverse charge distributions.

\begin{figure}[t]
  \centering
  \includegraphics[width=82mm]{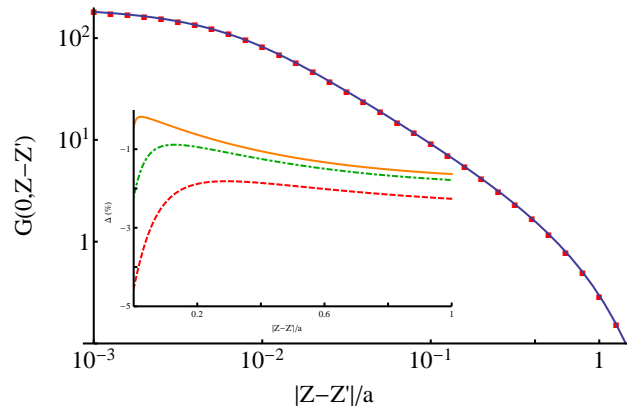}
  \caption{(Color online) The on-axis Green's function, red squares, and its
    analythical approximation, blue line, for the uniform transverse charge
    density with $b/a=0.01$. The inset shows the relative error, $\Delta$,
    between the exact and 
    analytical representations of the on-axis response function. The red dashed
    line corresponds to $b/a=0.1$, while the green dot-dashed and solid orange
    lines correspond to $b/a =0.05$ and $b/a=0.01$ respectively.}
  \label{fig:Fan}
\end{figure}

So far, we have found two analytical representations for the on-axis Green's
function that describe two physical limits. Figure~\ref{fig:Fs-Fi} compares
these approximate representations given by Eqs.~(\ref{eq:core_sum}) and
(\ref{eq:core_int}) with values of the on-axis Green's
function given in Eq.~(\ref{eq:on-axis-Greens}). It shows that approximate
expressions describe the on-axis Green's function well within their
applicability regions. We note that there is an overlap region where two
physical limits have common domain of mutual applicability, $b\ll |Z-Z'|\ll
a$. This region is well defined when the beam size is sufficiently smaller
than the pipe radius, $b < a$, and correspond to the case where the charge
distribution can be already approximated by a point charge  yet the presence of 
pipe walls can be still ignored. In this region, both expressions result in
the same asymptotic form for the on-axis Green's function and coincide with the
expression for the point charge potential in free space: 
\begin{equation}
  \label{eq:asymptot}
 G(0,Z-Z')=\frac{1}{|Z-Z'|},\ b\ll |Z-Z'|\ll a.
\end{equation}
This allows matching of the two analytical representations as their product divided by their
mutual asymptotic expression. As a result, the on-axis Green's function in a
perfectly conducting pipe can be approximated with: 
\begin{widetext}
\begin{equation}
  \label{eq:core}
G(0,Z-Z')\approx\frac{|Z-Z'|}{ a}
 \frac{\pi e^{-\mu _{0,1}|Z-Z'|/a}}{1-e^{-\pi |Z-Z'|/a}} \int d^{2}\mathbf{r'}
\frac{\rho_{\perp}(r')}{\sqrt{(Z-Z')^{2}+r'^{2}}} ,
\end{equation}
\end{widetext}
for $b< a$ and $ |Z-Z'|\in (0,\infty)$. 

\begin{table}[t]
\begin{ruledtabular}
  \begin{tabular}{l l l l}
    $\rho_{\perp}(r)$ & & $b$ & $\int d^{2}\mathbf{r'}
    \frac{\rho_{\perp}(r')}{\sqrt{(Z-Z')^{2}+r'^{2}}}$\\ 
    \hline

    $\frac{1}{\pi b^{2}},$ & $r\le b$ & $2x_{rms}$&
    $\frac{2}{b}\frac{1}{|\Delta Z| + \sqrt{\Delta Z^{2} + 1}} $ \\

    $\frac{2}{\pi b^{2}} \left(1-\frac{r^{2}}{b^{2}} \right),$ & $r\le b$ & $\sqrt{6} x_{rms}$&
    $\frac{8}{3b}\frac{1+ 3 \Delta Z^{2}/4}{|\Delta Z|^{3}+ \left( \Delta Z^{2}
          + 1\right)^{3/2}+ 3|\Delta Z|/2 }
      $ \\  

    $\frac{1}{\pi b^{2}}e^{-r^{2}/b^{2}},$ & $r\le \infty $& $\sqrt{2}x_{rms}$
    &
    $\frac{\sqrt{\pi}}{b} Erfc \left(|\Delta Z| \right) e^{\Delta Z^{2}}$ 
  \end{tabular}
\end{ruledtabular}
  \caption{The transverse density distributions with their rms sizes and the corresponding on-axis
    Green's functions in the vicinity of the sources with $\Delta Z =
    (Z-Z')/b$.}
  \label{tab:transverse-distributions}
\end{table}

The main advantage of using the compact form expression is its simplicity. In
contrast, evaluation of an exact Green's function described with
Eq.~(\ref{eq:on-axis-Greens}) requires summation of a large number of terms
and takes a significant amount of time. Figure~\ref{fig:Fan} shows the
comparison between these two approaches for the uniform transverse charge
distribution with $b/a=0.01$. The inset shows the relative error of using the
approximate Green's function instead of exact. The absolute error is
the largest at $Z=Z'$ but does not exceed $0.87069/a$ based on a simple method
of summing Bessel series proposed by Greenwood in
Ref.~\cite{a:Greenwood:1968}.

\section{\label{sec:IGF-approach} The Integrated Green's Function Approach} 

In the previous section, we have provided a compact form for the
Green's function that describes an axially symmetric space charge
problem. Using this result, the longitudinal space charge field takes on the
following form for the translationally invariant transverse distributions 
\begin{equation}
  \label{eq:Ez1}
E_{z}(r,z)=\frac{1}{4 \pi \varepsilon_{0} \gamma} \int _{-\infty }^{\infty}
\lambda(z')
\frac{\partial}{\partial z'} g_{z}(r, z-z') dz',
\end{equation}
with $g_{z}(r,z-z')=J_0\left(\frac{r}{\gamma} \frac{\partial}{\partial z} \right) g(0,z-
z') $. The operational form for the Green's function suggested in
Eq.~\ref{eq:op-response-func} also allows 
for a similar representation for the transverse component of the space charge
induced field 
\begin{equation}
  \label{eq:Er1}
 E_{r}(r,z)=\frac{1}{4 \pi \varepsilon_{0} } \int _{-\infty }^{\infty }
 \lambda(z') \frac{\partial}{\partial z'}  g_{r}(r,z-z')dz',  
\end{equation}
with $g_{r}(r,z-z')=-J_1\left(\frac{r}{\gamma} \frac{\partial}{\partial z}\right) g(0, z-
z')$ \footnote{Here we substituted
  $\frac{\partial}{\partial r} g(r,z-z')=- 
\frac{\partial}{\gamma\partial z} J_1\left(\frac{r}{\gamma} \frac{\partial}{\partial z} \right)
g(0,z-z') $ based on the formal definition of the function that contains a
derivative.}. This solution for the space charge problem allows for 
efficient numerical evaluation based on the Integrated Green's function
approach \cite{ip:Ryne:2003, ip:Abell:2007, a:Qiang:2007}. 

For the purposes of providing a recipe for numerical evaluation of the
Eqs.~\ref{eq:Ez1} and \ref{eq:Er1},  we will use the Integrated Green's
function approach in a constant function basis. This approach approximates the
charge density as $\lambda(z') = \left( \lambda_{j+1} +
  \lambda_{j} \right)/2$ within a cell of the computational domain, $z'\in \left \lbrack
  z_{j}, z_{j+1} \right \rbrack$ for $z_{j}=jh$, and arrives to the following discrete
convolution 
\begin{equation}
  \label{eq:IGF-step2}
  E_{r,z}(z_{i}) = h A_{r,z} \sum_{j} \lambda_{j}w_{r,z}^{i-j},
\end{equation}
with normalization constants $A_{r} = 1/4 \pi \varepsilon_{0}$ and
$A_{z}=A_{r}/\gamma$.

The Green's functions derived in this paper behave differently near the
source. Namely, the longitudinal Green's function, $g_{z}(r,z-z')$, is 
a continuous function of $z$, while the radial Green's function,
$g_{r}(r,z-z')$, has a discontinuity at $z=z'$. Taking this into account,
leads to the following integrated Green's function for the longitudinal
component
\begin{equation}
  \label{eq:zIGF}
w_{z}(\zeta)= \frac{g_{z}(r, \zeta-h ) - g_{z}(r, \zeta+h )}{ 2 h},
\end{equation}
with $\zeta = h(i-j)$; and for the radial component:
\begin{equation}
  \label{eq:rIGF}
w_{r}(\zeta) = \frac{ g_{r}(r, \zeta-h ) - g_{r}(r, \zeta+h )}{ 2 h} +
\frac{1}{h} g_{r}(r, 0^{-} )\delta_{\zeta,(-h,0,h)},
\end{equation}
where $w_{r}(h)$ uses the Green's function defined for $z\ge z'$ in order to
evaluate $g_{r}(r,0)$, while $w_{r}(-h)$ and $g_{r}(r, 0^{-} )$ use $z\le z'$ definition.

\section{\label{sec:IGF-implementation} Discussion}

Equation \ref{eq:core} is the main result of the paper that has allowed us to
express the space charge induced fields in terms of the Green's functions that
have simple analythical representation. Equations \ref{eq:Ez1} and
\ref{eq:Er1} with the corresponding Green's function are the next important
result of this paper. These expressions for the components of the field
can be efficiently evaluated using the Integrated Green's functions,
Eqs.~\ref{eq:zIGF} and \ref{eq:rIGF}. 

It is often a case that the transverse beam size is a smallest scale of the
problem and that only the space charge induced fields within the beam are of
interest. Thus, the particle accelerator codes, similar to Elegant, include
only the effects of the space charge induced fields at $r=0$. X-ray free
electron lasers often operate in a space charge dominated regime and a more
diligent treatment of the space charge induced fields is in order. In what follows
we will illustrate the steps required to obtain the off-axis behavior for the
components of the space charge induces field.

The radial component of the space charge induced field, evaluated on
$z_{i}=ih$ grid, is equal to $E_{r}(z_{i}) = h A_{r} \sum_{j}
\lambda_{j}w_{r}^{i-j}$, where $\lambda_{j}$ is a space charge distribution on
the same grid and $A_{r}=1/4 \pi \varepsilon_{0}$. Equation~\ref{eq:rIGF}
defines the integrated Green's function, $w_{r}(\zeta)$, in terms of
\begin{equation}
  \label{eq:Gr}
  g_{r}(r,z-z')=-\frac{r}{2}\frac{\partial}{\gamma\partial z} g(0,z- z'),
\end{equation}
where we have kept the leading term in the Taylor series expansion of
$J_{1}(x)$ for the purpose of illustration only. In order to evaluate the
derivation of the on-axis Green's function, we can apply the logic used
in deriving the analythical representation of the on-axis Green's function
itself and arrive to the following expression:
\begin{eqnarray}
 \label{eq:dGdz}
 \nonumber
\frac{\partial}{\partial Z} G(0,Z-Z')&\approx&\frac{|Z-Z'|^{3}}{ a}
 \left(\frac{\partial}{\partial Z}\frac{\pi e^{-\mu _{0,1}|Z-Z'|/a}}{1-e^{-\pi
       |Z-Z'|/a}} \right) \times\\
&\times &\int d^{2}\mathbf{r'}
\frac{\rho_{\perp}(r')}{\left\lbrack (Z-Z')^{2}+r'^{2}\right \rbrack^{3/2}}.
\end{eqnarray}

Evaluation of the longitudinal component of the space charge induced field
follows the same steps but uses $A_{z}=A_{r}/\gamma$ and $w_{z}(\zeta)$ as
defined in Eq.~\ref{eq:zIGF} in terms of 
\begin{equation}
  \label{eq:Gz}
  g_{z}(r,z-z')=\left(1-\frac{r^{2}}{4}\frac{\partial^{2}}{\gamma^{2}\partial z^{2}}
  \right) g(0,z - z'),
\end{equation}
where we have kept the first two leading terms in the Taylor series expansion of
$J_{0}(x)$ for the purpose of illustration only. The second term requires the
second derivative of the on-axis Green's function that has the following form:
\begin{eqnarray}
  \label{eq:d2Gdz2}
  \nonumber
\frac{\partial^{2}}{\partial Z^{2}} G(0,Z-Z')&\approx&\frac{|Z-Z'|^{3}}{2 a}
 \left(\frac{\partial^{2}}{\partial Z^{2}}\frac{\pi e^{-\mu _{0,1}|Z-Z'|/a}}{1-e^{-\pi
       |Z-Z'|/a}} \right)\times\\
&\times&        \int d^{2}\mathbf{r'}
\frac{\rho_{\perp}(r')\left\lbrack 2(Z-Z')^{2}-r'^{2}\right
  \rbrack}{\left\lbrack (Z-Z')^{2}+r'^{2}\right \rbrack^{5/2}}. 
\end{eqnarray}

From the provided illustration, the solution of the space charge problem at
$r=0$ requires a single evaluation of a discreet convolution. An additional
convolution provides a linear contribution to the radial component of the
space charge induced field. The third convolution can be also carried out in
order to find a next order correction to the longitudinal component of the
space charge induced field. One can apply the illustrated approach in order to
find the next order corrections with computational effort that scales linearly
with number of corrections.

\section{Conclusions}

We have studied the space charge problem for the beam with axially symmetric
charge distribution in a smooth perfectly conducting pipe. We have develop a
Green's function-based description for the space charge induced fields that
is valid in the full range of current variations and applicable to beams with
varying beam radius. The Green's function discussed in the paper corresponds
to the electrostatic potential of a charged disk in a pipe and is completely
defined by its behavior on the pipe axis due to axial symmetry. 

We have found a compact analytical approximation for the on-axis Green's
function that allows for analytical calculation of the Green's 
function off axis. Having a compact representation of the on-axis behavior
of the response function can improve semi-analytical codes, such as Elegant, as
it offers a significant advantage over numerical evaluation of the exact
solution.

Finally, we have provided a detail prescription based on the Integrated
Green's function approach for efficient numerical evaluation of
the fields in the case of translational symmetry. This approach describes
transverse as well as longitudinal component of the space charge induced field
and scales linearly with the number of radial corrections.

\section{Acknowledgements}
We gratefully acknowledge the support of the U.S.~Department of Energy through
the LANL/LDRD Program for this work. 

\bibliography{SpCharge}

\begin{thebibliography}{10}

\bibitem{ip:Abell:2007}
D.~T. Abell, P.~J. Mullowney, K.~Paul, V.~H. Ronjbar, and R.~D. Ryne.
\newblock In {\em Particle Accelerator Conference}, 2007.

\bibitem{tr:OPAL:2008}
A.~Adelmann, Ch. Kraus, Y.~Ineichen, S.~Russell, Yuanjie Bi, and J.~Yang.
\newblock The opal (object oriented parallel accelerator library) framework.
\newblock Technical Report PSI-PR-08-02, Paul Scherrer Institut, 2008.

\bibitem{tr:Billen:1996}
J.~Billen.
\newblock Parmela.
\newblock Technical Report LA-UR-96-1835, Los Alamos National Laboratory, 1996.

\bibitem{tr:Borland:2000}
M.~Borland.
\newblock Technical Report LS-287, ANL Advanced Photon Source, 2000.

\bibitem{a:Borland:2002}
M.~Borland, Y.C. Chae, P.~Emma, J.W. Lewellen, V.~Bharadwaj, W.M. Fawley,
  P.~Krejcik, C.~Limborg, S.V. Milton, H.-D. Nuhn, R.~Soliday, and M.~Woodley.
\newblock Start-to-end simulation of self-amplified spontaneous emission free
  electron lasers from the gun through the undulator.
\newblock {\em Nucl. Instr. Meth. Phys. Res. A}, 483(1–2):268 -- 272, 2002.
\newblock Proceedings of the 23rd International Free Electron Laser Conference
  and 8th FEL Users Workshop.

\bibitem{a:Bouwkamp:1947}
C.~J. Bouwkamp and N.~G. De~Bruijn.
\newblock The electrostatic field of a point charge inside a cylinder, in
  connection with wave guide theory.
\newblock {\em J. Appl. Phys.}, 18(6):562--577, 1947.

\bibitem{b:Chao:1993}
A.~Chao.
\newblock {\em Physics of Collective Beam Instabilities in High Energy
  Accelerators}.
\newblock Wiley, New York, 1993.

\bibitem{a:Chasman:1969}
R.~Chasman.
\newblock Numerical calculations on transverse emittance growth in bright linac
  beams.
\newblock {\em IEEE Trans. Nucl. Sci.}, 16(3):202--206, June 1969.

\bibitem{web:Floettmann:2011}
K~Floettmann.
\newblock Astra: A space charge tracking algorithm, 2011.

\bibitem{a:Greenwood:1968}
J.~A. Greenwood.
\newblock {\em Proc. Camb. Phil. Soc.}, 64:705--710, 1968.

\bibitem{a:Hofmann:1981}
I.~Hofmann.
\newblock Emittance growth of beams close to the space charge limit.
\newblock {\em IEEE Trans. Nucl. Sci.}, 28(3):2399--2401, June 1981.

\bibitem{a:Huang:2004}
Z.~Huang, M.~Borland, P.~J. Emma, J.~Wu, C.~Limborg, G.~Stupakov, and J.~Welch.
\newblock Suppression of microbunching instability in the linac coherent light
  source.
\newblock {\em Phys. Rev. ST Accel. Beams}, 7:074401, 2004.

\bibitem{a:Kim:1989}
K.J. Kim.
\newblock Rf and space-charge effects in laser-driven rf electron guns.
\newblock {\em Nucl. Instr. Meth. Phys. Res. A}, 275(2):201--218, 1989.

\bibitem{a:Kur:2011}
E.~Kur, D.~J. Dunning, B.~W.~J. McNeil, J.~Wurtele, and A.~A. Zholents.
\newblock A wide bandwidth free-electron laser with mode locking using current
  modulation.
\newblock {\em New J. Phys}, 13, 2011.

\bibitem{a:Lapostolle:1971}
P.M. Lapostolle.
\newblock Possible emittance increase through filamentation due to space charge
  in continuous beams.
\newblock {\em IEEE Trans. Nucl. Sci.}, 18(3):1101--1104, June 1971.

\bibitem{Note1}
Consider the following on axis potential, $G(0,Z-Z')=1/|Z-Z'|$. Using the
  Poisson representation, one can show that $G(R,Z-Z')=1/\protect \sqrt
  {(Z-Z')^{2}+R^{2}}$, which is a free space potential of a point charge in
  cylindrical coordinates. In order to apply our approach, one start with the
  Taylor series expansion, $J_{0}(x) = 1 - \protect \frac {x^{2}}{4} +
  O(x^{4})$, and arrives to $G(R,Z-Z') = 1/|Z-Z'| -\protect \frac {R^{2}}{4}
  \protect \frac {\partial ^{2}} {\partial Z^{2}} 1/|Z-Z'| + O(R^{4})$.
  Carrying out the derivatives leads to $G(R,Z-Z') = 1/|Z-Z'| \left (1
  -\protect \frac {R^{2}}{2(Z-Z')^{2}} + O(R^{4})\right ) $, which is a series
  expansion of a free space potential of a point charge in cylindrical
  coordinates for small $R$.

\bibitem{Note2}
Here we substituted $\protect \frac {\partial }{\partial r} g(r,z-z')=-
  \protect \frac {\partial }{\gamma \partial z} J_1\left (\protect \frac
  {r}{\gamma } \protect \frac {\partial }{\partial z} \right ) g(0,z-z') $
  based on the formal definition of the function that contains a derivative.

\bibitem{a:Poisson:1823}
S.~D. Poisson.
\newblock {\em Journal de l'Ecole Royale Polytechinque}, 12(19):215--248, 1823.

\bibitem{a:Qiang:2007}
J.~Qiang, S.~Lidia, R.~D. Ryne, and C.~Limborg-Deprey.
\newblock Three-dimensional quasistatic model for high brightnees beam dynamics
  simulation.
\newblock {\em Phys. Rev. ST Accel. Beams}, 10:129901, 2007.

\bibitem{tr:Rosenzweig:1996}
J.~Rosenzweig, C.~Pellegrini, L.~Seranfini, C.~Ternienden, and G.~Travish.
\newblock Space-charge oscillations in a self-modulated electron beam in
  multi-undulator free-electron lasers.
\newblock Technical Report TESLA-FEL-96-15, DESY, 1996.

\bibitem{ip:Ryne:2003}
R.~D. Ryne.
\newblock In {\em ICFA Beam Dinaymcs Workshop on Space-Charge Simulation},
  Oxford, 2003.

\bibitem{a:Tanaka:2013}
Takashi Tanaka.
\newblock Proposal for a pulse-compression scheme in x-ray free-electron lasers
  to generate a multiterawatt, attosecond x-ray pulse.
\newblock {\em Phys. Rev. Lett.}, 110(8), FEB 20 2013.

\bibitem{a:Ventury:2008}
M.~Venturini.
\newblock Models of longitudinal space-charge impedance for microbunching
  instability.
\newblock {\em Phys. Rev. ST Accel. Beams}, 11:034401, 2008.

\bibitem{a:Zholents:2005}
A.~A. Zholents.
\newblock Method of an enhanced self-amplified spontaneous emission for x-ray
  free electron lasers.
\newblock {\em Phys. Rev. ST Accel. Beams}, 8(4), APR 2005.

\end{thebibliography}

\appendix
\section{Green's function}

The Green's function for the Laplace's equation inside a perfectly conducting
pipe must satisfy the following equation
\begin{equation}
  \label{eq:Greens-equation}
  \nabla^{2}\Gamma(\mathbf{R}|\mathbf{R'})=-4\pi \delta(\mathbf{R}-\mathbf{R'})
\end{equation}
with the boundary condition $\Gamma(R=a)=0$ and $\Gamma(Z\to\pm\infty)=0$.

To find a Green's function, it is common to expand the Green's function in
terms of axial eigenfunction solutions of the Laplace's problem in cylindrical
coordinates, $e^{i k Z}$. Here, however, the radial eigenfunction expansion is
used in order to allow for beam size variation along the beam. Thus 
\begin{equation}
  \label{eq:Greens-expansion}
  \Gamma( \mathbf{R} | \mathbf{R'}) = \sum_{m=-\infty}^{\infty}
  \sum_{n=1}^{\infty} G_{m,n}(Z|\mathbf{R'}) J_{m}(\mu_{m,n}R/a) e^{i m\Phi}, 
\end{equation}
where $J_{m}$ is the ordinary Bessel's function of order $m$ and $\mu _{m,n}$
is defined as a solution of $J_m\left(\mu _{m,n}\right)=0$. Substituting this
expansion in \ref{eq:Greens-equation} we find that
\begin{eqnarray}
  \label{eq:Greens-equation-expansion}
  \nonumber
&&  \nabla^{2}\Gamma(\mathbf{R}|\mathbf{R'})=\sum_{m=-\infty}^{\infty}
  \sum_{n=1}^{\infty} \left\lbrack \frac{d^{2}G_{m,n}(Z|\mathbf{R'})} {dZ^{2}} -\right.\\
&&\left.- \frac{\mu_{m,n}^{2}}
    {a^{2}} G_{m,n} (Z|\mathbf{R'})\right\rbrack J_{m}(\mu_{m,n}R/a)e^{i m \Phi}. 
\end{eqnarray}

With the help of the orthogonality properties for the Bessel's function, we
can now obtain that
\begin{eqnarray}
  \label{eq:Gm-equation}
  \nonumber
&&  \frac{d^{2}G_{m,n}(Z|\mathbf{R'})} {dZ^{2}} - \frac{\mu_{m,n}^{2}} {a^{2}} G_{m,n}(Z|\mathbf{R'}) =\\
&&  -\frac{4} {a^{2} J_{m+1}(\mu_{m,n})^{2}} J_{m}(\mu_{m,n} R'/a) 
  e^{-i m \Phi'} \delta(Z-Z').
\end{eqnarray}

Then, if we look at the points where $z\ne z'$, (\ref{eq:Gm-equation}) becomes
a simple eigen function problem with a well known solution
\begin{equation}
  \label{eq:exponential-solution}
  G_{m,n}(Z|\mathbf{R'})= g_{m,n}(R',\Phi')e^{-\mu_{m,n}|Z-Z'|/a}.
\end{equation}
The condition for the discontinuity of the first derivative at $Z=Z'$ yields that
\begin{equation}
  g_{m,n}(R',\Phi')=
      \frac{2}{a \mu_{m,n}J_{m+1}(\mu_{m,n})^{2}} J_{m}(\mu_{m,n}R'/a) e^{-i m
        \Phi'} .
\end{equation}

Finally, the Green's function for axially symmetric charge distributions is
\begin{eqnarray}
  \label{eq:Greens-symmetric}
&&  \Gamma(R,Z|R',Z')=\sum_{n=1}^{\infty} \frac{2}{a
    \mu_{0,n}J_{1}(\mu_{0,n})^{2}}\times\\
&&\times  J_{0}(\mu_{0,n}R/a) J_{0}(\mu_{0,n}R'/a)  e^{-\mu_{0,n}|Z-Z'|/a},
\end{eqnarray}

\section{\label{sec:comparison} Comparison with previous results}

Let us apply the results of the paper for analytical analysis of the
space charge problem. In what follows, we will use Eq.~\ref{eq:Ez1} in order
to derive a previously known expressions for the longitudinal component of the
charge induced electric field in long- \cite{b:Chao:1993} and short-scale
\cite{tr:Rosenzweig:1996} current variation limits.

The short-scale current variations commonly arise due to microbunching
instability that is driven by the longitudinal component of the space charge
induced electric field. It is common to assume a beam with a circular cross
section of radius $b$ and a constant transverse density profile. In the case
that observation point is located on-axis ($r=0$), one defines an impedance
(per unit length) $Z(k)$ as: 
\begin{equation}
  \label{eq:Ez-short}
  E_{z}(k)=-Z(k)I(k) , 
\end{equation}
where $I(k)$ is the Fourier component of the current $I(z)=c \lambda(z)$, with
$\lambda(k)=(2\pi)^{-1}\int_{-\infty}^{\infty}\lambda(z) e^{-i k z}dz$. 
 The impedance that has been
implemented in Elegant \cite{tr:Borland:2000} to simulate space charge effect
on a beam in a drift space \cite{a:Huang:2004} is
\begin{equation}
  \label{eq:Z0}
Z(k)=\frac{i}{\pi\varepsilon_{0}c k b^{2}}
\left[1-\frac{k b}{\gamma}  K_1\left(\frac{ k b}{\gamma} \right)\right],
\end{equation}
where $K_{1}$ is the modified Bessel function of the second kind
\cite{tr:Rosenzweig:1996}.

The longitudinal space charge impedance in Eq.~\ref{eq:Z0} is valid in the
short-wavelength limit, $k\to \infty $, where the effect of boundaries can be
neglected. The impedance that includes the effect of boundaries has the
following long-wavelength limit: 
\begin{equation}
  \label{eq:Zlong}
Z(k\to 0)=\frac{i}{4\pi\varepsilon_{0}c}\frac{k}{\gamma^{2}} \left[ 1 - 2 \log \left(
    \frac{b}{a} \right) \right],
\end{equation}
according to Ref.~\cite{a:Ventury:2008}. This impedance corresponds to the
on-axis case of another well-known expression for the longitudinal component
of the space charge induced electric field \cite{b:Chao:1993}: 
\begin{equation}
  \label{eq:chao}
E_{z}(r,z) \approx -\frac{1}{4\pi\varepsilon_{0}c \gamma^{2}} \frac{ d
  I(z)} {d z}\left[ 1 -\frac{r^{2}}{b^{2}}- 2 \log \left(
    \frac{b}{a} \right) \right].
\end{equation}

\section{\label{sec:lr-limit} Long-scale current variation limit}

We begin our comparison with a long-scale current variation limit. According
to Eq.~\ref{eq:chao}, the longitudinal component of the space charge induced
electric field is proportional to the current derivative. Integrating
Eq.~\ref{eq:Ez1} by parts and assuming that charge derivative, $d \lambda(z) /
dz = c^{-1} dI(z) / dz$, is constant on a scale $\delta z\gg a/\gamma$, we
obtain the following expression for the longitudinal component of the space
charge induced electric field: 
\begin{equation}
  \label{eq:Ez-long}
  E_{z}(r,z)=-\frac{1}{4 \pi \varepsilon_{0} c\gamma^{2}} \frac{d I(z)}{d
    z}\int _{-\infty }^{\infty}  G(r, Z') dZ',
\end{equation}
where the Green's function is $G(r, Z') = G(0,Z')
-\frac{r^{2}}{4}\frac{\partial^{2}}{\partial Z'^{2}} G(0,Z')$.

The $r^{2}$-term for the longitudinal component of the space charge induced
field is proportional to the integral of the second derivative of the on-axis
Green's function, which is equal to $-2\frac{\partial}{\partial Z'} G(0,Z'
\to 0^{+})$. Let us recall here that the on-axis Green's function is the
electrostatic potential of a charge disk and thus its negative derivative is
equal to electric field. Using $E_{Z}(0,z\to 0^{+})=2\pi \rho_{\perp}(0)$ for the
uniform transverse charge distribution, one obtains that the value of
coefficient is $4/b^{2}$.

The integral of the on-axis Green's function for the uniform transverse charge
distribution is
\begin{equation}
  \int _{-\infty }^{\infty}  G(0, Z') dz' = \frac{1}{x}\sum_{n=1}^{\infty}
  \frac{8} {\mu_{0,n}^{3}J_{1}(\mu_{0,n})^{2}}  J_{1}\left(\mu_{0,n} x \right),
\end{equation}
with $x=b/a$.
Thus, in order to show the equivalence with Eq.~\ref{eq:chao}, one
has to show that 
\begin{equation}
\label{eq:math}
  x(1-2\log x) = \sum_{n=1}^{\infty} \frac{8
    }{
    \mu_{0,n}^{3}J_{1}(\mu_{0,n})^{2}}  J_{1}\left(\mu_{0,n} x \right),
\end{equation}
which is the case of Dini expansion of the function.

The Dini expansion of a function defined on the interval  $x\in [0,1]$ with
the following boundary condition $f(1)+f'(1)=0$ is
$f(x)=\sum_{n=1}^{\infty}c_{n}J_{1}(\mu_{0,n}x)$ with the coefficients
\begin{equation}
c_n=\frac{2}{J_1\left(\mu _{0,n}\right)^2}\int _0^1x f(x)J_1\left(\mu
  _{0,n}x\right)dx. 
\end{equation}
Based on recursion relations $J_{1}(x)=-dJ_{0}(x)/dx$ and
$J_{0}(x)=x^{-1}d[xJ_{1}(x)]/dx$, one can show that the expansion coefficients
of $f(x)=x\left(1-2\log x\right)$ are indeed equal to
$c_{n}=8/\mu_{0,n}^{3}J_{1}(\mu_{0,n})^{2}$ and prove Eq.~(\ref{eq:math}).
Thus, we conclude that the Green's function based approach reproduces the
well-known result for the longitudinal component of the space charge induced
field in the limit of a long-scale current variation and constant beam radius
\cite{b:Chao:1993}. 

\section{\label{sec:sr-limit}Short-range current variation}

The Green's function description, presented in this paper provides an
alternative expression for the longitudinal space charge impedance:
\begin{equation}
  \label{eq:Z-integral}
Z(k)=\frac{i k }{c\gamma^{2}} \int _{-\infty }^{\infty}
e^{-i k z'/\gamma} \Phi\left(0,0,z'\right) dz',
\end{equation}
and thus is determined by the Fourier spectrum of the on-axis response
function. Evaluation of the Fourier spectrum of the on-axis response function
according to  Eq.~(\ref{eq:op-response-func}), leads to the following
representation of the longitudinal space charge impedance: 
\begin{equation}
  \label{eq:Z-const}
\hat{Z}_{\text{lsc}}(k)= \frac{i}{\pi\varepsilon_{0}c k b^{2}} \left[ \frac{b}{a}
  \sum_{n=1}^{\infty }\frac{ A_{n}}{1+\mu_{0,n}^{2}\gamma^{2}/k^{2}a^{2}}
  J_1\left(\mu_{0,n}\frac{b}{a}\right)\right],
\end{equation}
where the sum is due to the presence of the pipe. 

The free space approximation used to describe the limit of short-range current
variations ignores the presence of the pipe. Thus it replaces the discrete
modes spectrum inside the pipe with the continuous spectrum of the free
space. This reflects the fact that a large number of transverse modes
contribute to the space charge filed. In this case consecutive terms in the
sum in Eq.~(\ref{eq:Z-const}) are close to each other and the sum can be
replaced with the integral. Transition from the sum to an integral over
$x=b\mu_{0,n}/a$ has the following Jacobian $dn/dx=a/\pi b$. The resulting
integral on an interval $x\in \left[0,\infty\right]$ is known and leads to the
expression given by Eq.~(\ref{eq:Z0}).

\end{document}